# Natural Spider Silk Nanofibrils Produced by Assembling Molecules or Disassembling Fibers

Dinidu Perera,† Linxuan Li,† Chloe Walsh, Qijue Wang, Hannes C. Schniepp*

**Affiliations**
   Applied Science Department, William & Mary, P.O. Box 8795, Williamsburg, VA, 23187-8795, USA.

* Corresponding author. Email: schniepp@wm.edu. Phone: +1 757-221-2559. Fax: +1 757-221-2050
† These authors contributed equally to this work.

**Abstract**

Spider silk is biocompatible, biodegradable, and rivals some of the best synthetic materials in terms of strength and toughness. Despite extensive research, comprehensive experimental evidence of the formation and morphology of its internal structure is still limited and controversially discussed. Here, we report the complete mechanical decomposition of natural silk fibers from the golden silk orb-weaver *Trichonephila clavipes* into ≈10 nm-diameter nanofibrils, the material's apparent fundamental building blocks. Furthermore, we produced nanofibrils of virtually identical morphology by triggering an intrinsic self-assembly mechanism of the silk proteins. Independent physico-chemical fibrillation triggers were revealed, enabling fiber assembly from stored precursors "at-will". This knowledge furthers the understanding of this exceptional material's fundamentals, and ultimately, leads toward the realization of silk-based high-performance materials.

## 1. Introduction

    Spider silk features an impressive tensile strength and toughness, outperforming many man-made engineering materials [1–5]. This high-performance protein-based material is fully sustainable and produced under ambient temperature and pressure, using a small amount of energy [4]. Furthermore, its biocompatibility makes it attractive for biomedical applications. Because of these outstanding traits and prospects, spider silk has been intensely studied. Yet, details of the structure and formation process of the impressive high-performance silk fibers are

still not understood well enough [6–8] to enable the synthesis of silk-inspired fibers with comparable properties [7,9–11]. Here, we demonstrate that the native silk protein has a pronounced preference to form nanofibrils, about 10 nm in diameter: we have found these nanofibrils to represent the bulk of naturally produced spider silk fibers, and we have found that they can be triggered to naturally self-assemble from native silk dope *in vitro*. This has wide-ranging implications for the understanding of the structure of spider silk fibers, their structure–property relationships, and their synthesis.

Nanofibrils have been shown to be the basic building blocks of many strong and hard biological materials such as exoskeletons of crustaceans (chitin), wood (cellulose), tissues/tendons/cartilage (collagen), as well as the silk of silkworms [12,13]. In contrast, the role and prevalence of nanofibrils within spider silk, a material providing a unique combination of high strength and high extensibility, has not yet been fully established. While several widely used models suggest silk nanofibrils as the predominant structure in the core of the fibers, there is hardly any experimental evidence demonstrating that they are indeed present in spider silk fibers in large amounts [1,2,6,14–16]. As a matter of fact, some studies still promote globular proteins as basic structural elements of spider silk [17,18]. Moreover, studies detecting nanofibrils in spider silk feature widespread regarding their lengths and diameters between different species or even within a single species [6,19–22]. We recently discovered that the tape-like silk of the Chilean recluse spider [23] is solely made out of 20 nm-diameter nanofibrils, which are oriented strictly parallel to the direction of the fiber [24]. While this observation provides important support for the idea that nanofibrils are pivotal for the structure and performance of spider silk, it is not clear whether the results of this relatively uncommon spider silk readily apply to most or all spider silks. The most commonly studied species are orb weavers featuring cylindrical fiber morphologies, in particular the *Trichonephila*, *Araneus*, or *Argiope* genera, which have become key model systems [3].

The exfoliation of silk fibers, predominantly from silkworms, has recently been studied and demonstrated by several groups [25–31]. This technique offers the prospect of reshaping silk into other form factors without losing some of its key traits, and it is thus interesting for many applications, such as making papers [30,31], filter membranes [27], sensors, and electronic devices [25,26]. In addition, these techniques can also provide an insight into the internal structural makeup of the fiber. Achieving a high degree of exfoliation without excessive protein degradation is challenging; chemical and mechanical treatments and combinations thereof have been used. The chemical treatments have included acid hydrolysis [32], high concentrations of urea [33], deep eutectic solvents (guanidine hydrochloride and urea) [27], hexafluoroisopropanol (HFIP) [34], and NaOH with urea [35]. As mechanical means, ultrasonication and high-shear mixing have been employed [28,30,36]. Most of these efforts have yielded nano- or microfibrillar materials; however, depending on the experimental techniques used, a widespread in terms of the degree of exfoliation, fibril diameters, and diameter distributions have been reported, ranging from 3 nm to hundreds of nanometers [6,12]. For spider silk, on the other hand, only two studies have attempted exfoliation, without providing information on the exfoliation yield or the role of nanofibrils within the silk fiber [36,37]. In particular, exfoliation has not been reported for the important silk of orb-weaving spiders.

Studying exfoliation can reveal the prevalence and properties of nanofibrils within a silk fiber, which is important knowledge for the design of synthetic, silk-inspired fibers. However, for the synthesis of such fibers, it is equally important to know how these fibrils come into existence, starting from the aqueous silk dope, and how they become oriented within the fiber. Interestingly, we have been able to demonstrate that native silkworm silk proteins (fibroins) taken directly from the gland can be triggered to self-assemble into long, straight ≈20 nm nanofibrils closely resembling the nanofibrils observed in natural or exfoliated silkworm silk by applying shear [38]. This capability was absent in reconstituted silk proteins [39]. Similar types of self-assembly have also been studied with recombinant spider silk proteins (spidroins), even to synthesize macroscopic fibers [7,40–42]. However, these recombinant spidroins currently do not implement the full sequence of their natural counterparts, and the materials made from these recombinant spidroins typically exhibit lower performance than natural spider silk [7–9,40]. Recently, the early, pre-fibrillar stages of self-assembly from hierarchical micellar subdomains induced by shear have been studied in native *Latrodectus hesperus* spidroin by nuclear magnetic resonance and cryo-transmission electron microscopy [43]. However, the self-assembly of native spidroin nanofibrils has not been reported yet.

To produce protein fibers that can be spontaneously produced from a protein reservoir as in the natural system, one needs not only an effective self-assembly mechanism to produce nanofibrils with high yield, but also a way to prevent premature assembly of the stored protein. To prevent premature assembly inside the storage sack, silk protein is stored in an environment with neutral pH and high $Na^+$ and $Cl^-$ ion concentrations [1,44–49]. It has been proposed that in these conditions, spidroins form micellar-like structures to prevent aggregation and unwanted fibrillation. When the silk dope travels through the S-shaped tapering duct, the pH reduces, and $Na^+$ and $Cl^-$ ions exchange with $K^+$ and $PO_4^{3-}$ ions [1,44–49]. These chemical changes with the help of physical shearing force generated by the tapering duct trigger the self-assembly of aligned nanofibrils from the silk dope.

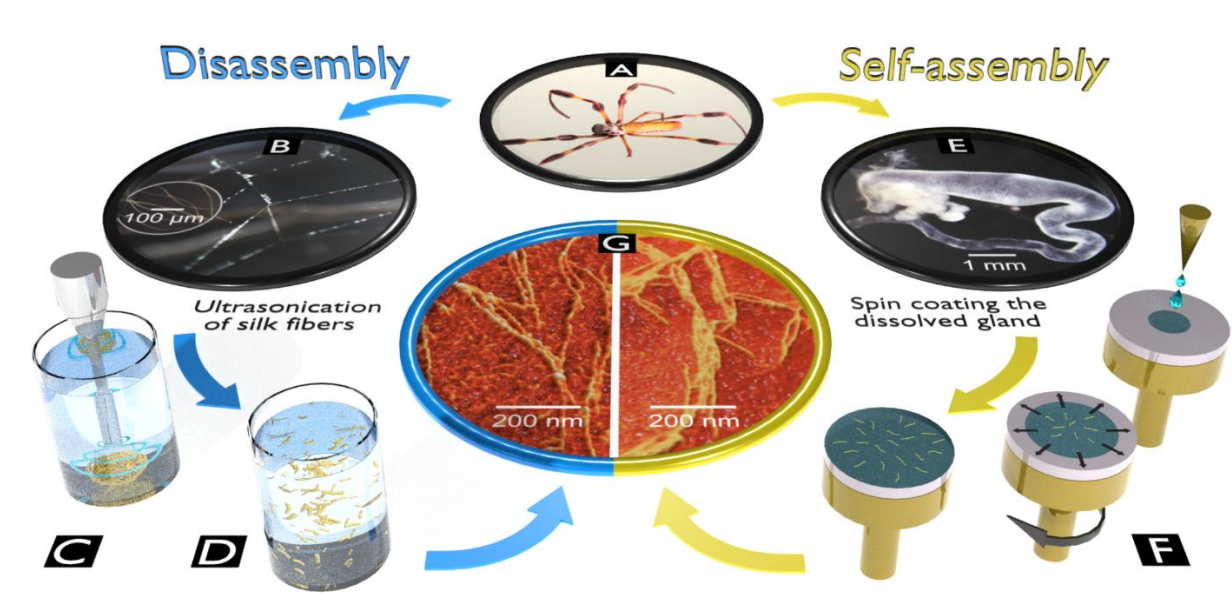

**Fig. 1. Spider silk nanofibrils via disassembly or assembly.** (**A**) Live *Trichonephila clavipes* spider. (**B**) Web naturally spun by *T. clavipes*. (**C**), (**D**) Ultrasonication (**C**) of collected silk fibers and subsequent exfoliation into nanofibrils (**D**). (**E**) Major ampullate silk gland from dissected *T. clavipes* spider (optical micrograph). (**F**) Spin coating the obtained, diluted silk dope for nanofibril self-assembly. (**G**) AFM topography of exfoliated (left) and self-assembled (right) silk nanofibrils.

Here we report the preparation of spider silk nanofibrils using both disassembly of native fibers via exfoliation and self-assembly of molecules from native silk dope (Fig. 1). In both cases, the silk was sourced from *Trichonephila clavipes* (Fig. 1A), arguably the most well-studied system for mechanically high-performing silks. For exfoliation, we collected major ampullate (MA) silk naturally spun by spiders living in large cages (Fig. 1B). Exfoliation was done using an ultrasonic homogenizer (Fig. 1C) until the macroscopic silk fibers disintegrated (Fig. 1D). For self-assembly, we dissected live spiders to extract the major ampullate silk gland (Fig. 1E) and subsequently diluted it using deionized water to prepare solutions with a range of concentrations. These solutions were spin-coated onto dry mica substrates, a process that shears and concentrates the silk solutions within seconds, not unlike what happens in the spider's spinning duct during the natural spinning process (Fig. 1F). Interestingly, dynamic-mode atomic force microscopy (AFM) scans of these samples revealed that both of these approaches — exfoliation and self-assembly — yielded uniform nanofibrils featuring similar morphologies and similar diameters of ≈10 nm (Fig. 1G). Our findings suggest that the formation of spider silk nanofibrils from an aqueous solution of spidroin is a robust process and an intrinsic capability of these proteins. Notably, we also observed that even in the presence of shear force, this self-assembly process could be prevented by maintaining a solution chemistry similar to what is found in the storage sack: a close to neutral pH and a high concentration of $Na^+$ and $Cl^-$ ions. These findings show for the first time that physical and chemical conditions are both critical controlling factors for the nanofibril formation from native spidroin. This insight into the spinning conditions is particularly important because nanofibrils have been shown to play a decisive role in the outstanding structural properties of silk by experimental and theoretical studies. Thus, our work provides significant inspiration for ongoing efforts to synthesize artificial yet sustainable silk-like fibers with strength and toughness comparable to the natural material.

## 2. Materials and Methods

### 2.1 Sample Preparation

Mature female *Trichonephila clavipes* spiders were captured in the wild in Citrus County, Florida. They were kept in a cage made of Plexiglass strips and aluminum mesh and were fed one cricket per week.

For exfoliation experiments, naturally spun silk was collected from the cages of healthy-looking spiders; silk (3.4 mg) was added to deionized water (60 mL) (Millipore Synergy UV). Ultrasonication (750 W, 20 kHz, 40% amplitude) was applied to the solution (Cole-Parmer CPX750 Ultrasonic Homogenizer) at a controlled temperature of 50 °C. Ultrasonic pulses of 10 seconds duration followed by a 10-second pause were run for 30–135 minutes to obtain the desired level of exfoliation. Then, 40 μL of the resulting solution was sampled and spin-casted (Laurell WS-400Bz-6NPP) onto freshly cleaved mica sheets (2 mins at 2000 rpm) for imaging.

For self-assembly experiments, the major ampullate glands were obtained as described by Jeffery et al.[50] Then, forceps were utilized to poke a hole into the glycoprotein layer of the gland to expose the silk dope inside. The dope was then placed in a microcentrifuge tube containing water (1 mL) (or PBS buffer for the experiments we did to analyze the effect of ions) and left to homogenize overnight in a refrigerator to prevent the proteins in the dope from denaturing. To determine the original concentration of the stock solution, a drop of known volume $V_s$ from the solution was placed on a substrate of mass $m_s$, which was then placed in a vacuum oven at 60 °C to dry overnight, and the mass $m_l$ of the dried sample was measured. The concentration $c$ of the stock solution was then calculated as $c = (m_l - m_s)/V_s$. Dilutions of 1000 mg/L, 100 mg/L, and 10 mg/L were then prepared from these stock solutions. 40 µL of solution were micro-pipetted onto freshly cleaved mica and spin-coated (Laurell WS-400Bz-6NPP) for 2 minutes at 2000 rpm.

## 2.2 Atomic Force Microscopy

NTEGRA Prima Scanning Probe Laboratory (NT-MDT, Zelenograd, Russia) was utilized to scan both exfoliated and self-assembled samples at room temperature. To minimize the deformation caused by tip-sample interactions, dynamic mode imaging was conducted using µmasch HQ: NSC15/Al BS AFM tips featuring a radius of curvature $r \approx 8$ nm, a resonance frequency $f \approx 325$ kHz, and a spring constant $k \approx 40$ Nm$^{-1}$. Gwyddion software was used to flatten and analyze the AFM topography images (http://gwyddion.net/). Blender software (https://www.blender.org/) was used to prepare 3D topography images and 3D rendered illustrations of the proposed silk structure.

## 2.3 Scanning Electron Microscopy

A Hitachi S-4700 field emission electron microscope (FESEM) equipped with a secondary electron detector was used to image disintegrated silk fibers. The samples were run at an acceleration voltage of 10 kV with a 7.5–8.0 mm working distance. For sample preparation, disintegrated silk fibers were dried on a mica sheet and attached to an aluminum SEM sample holder via conductive tape. The sample was sputter-coated (Anatech LTD, Hummer 6.2) with ≈2 nm layer of gold/palladium prior to FESEM imaging to prevent charging.

## 2.4 Statistical Analysis

The diameter measurements of fibers in SEM images were done using the ImageJ image analysis software (https://imagej.nih.gov/ij/). The diameters of nanofibrils in AFM images were measured using the following procedure. Topography sections perpendicular to nanofibrils were first produced from AFM topography data using the Gwyddion AFM image processing software. The topography profiles of the fibrils were then fitted with Gaussian functions, and the full width at half-maximum values of these fitted curves were taken as the diameters of nanofibrils. This criterion was used to compensate for the broadening induced by the size of the AFM probe, which is comparable to the fibril diameters. All data analysis was implemented in MATLAB (*MATLAB ver. R2019a*). The raw data and codes, along with explanations of all methods used for data analysis, have been published in a data repository [51]. All values were expressed as mean ± standard deviation, and no further statistical analysis was used.

## 3. Results and Discussion

### 3.1 Exfoliation reveals 10 nm diameter nanofibrils as fundamental building blocks

We employed ultrasonication, a purely mechanical method, to disintegrate *T. clavipes* silk in deionized water. Notably, this approach avoids harsh or hazardous chemicals that degrade the protein structure and compromise the biocompatibility of silk [25,52]. After 30 minutes of ultrasonication, the solution was cloudy, and optical microscopy revealed that the macroscopic fibers had begun exfoliating (Fig. 2, A and B). Scanning electron microscopy (SEM, Fig. 2D) of the exfoliated material further revealed the fibers' internal fibrillar structure. Toward the end of each fiber fragment (Fig. 2C), exfoliation had progressed all the way to the core, eventually compromising the integrity of the fiber as a whole. The process decomposed the fiber into thinner and thinner fibrils (Fig. 2F), successively decreasing diameters from 300 to 40 nm (Fig. 2, G and H). No particular fibril diameter was dominant, especially when volume averages were considered (Fig. 2H). The latter were obtained by weighting the fibril counts with the square of the fibril diameter, representing the cross-sectional area of the fibrils. This observation indicates the absence of nanofibril bundles of a particular diameter, contrary to what has been suggested in the literature [13,53,54].

To observe material that was completely removed from the fiber in this process and ended up in solution, we spin-coated a small droplet of the solution onto a mica substrate and imaged it using dynamic-mode AFM. Interestingly, the AFM images showed a homogeneous population of nanofibrils with diameters of ≈10 nm (Fig. 2E). The fact that we do not observe the smallest, 10 nm-diameter nanofibrils in our SEM images is most likely due to the resolution limitations of our hardware. Furthermore, images taken on dried samples may underestimate the degree of exfoliation (and perhaps overestimate the fibril diameters) due to the re-aggregation and re-assembly of already exfoliated fibrils in the drying process. Importantly, when we increased the ultrasonication duration to 135 minutes, virtually no fiber fragments were detectable in the solution using optical microscopy. AFM analysis after spin coating this solution onto mica (Fig. 3A) revealed nanofibrils similar to Fig. 2E. Therefore, we conclude that this longer ultrasonication fully exfoliated the 3.4 mg of silk into a solution of nanofibrils with a concentration of 0.09 mg/mL.

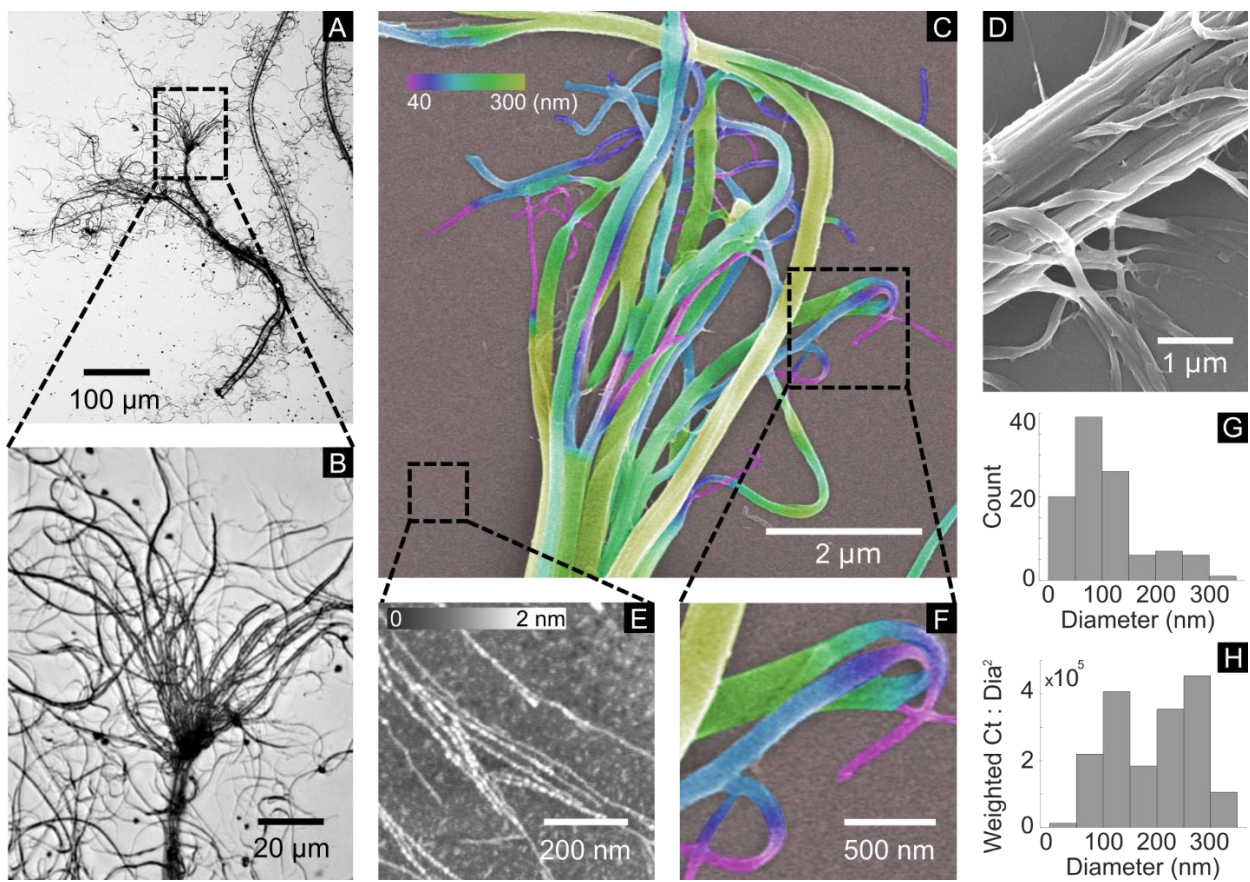

**Fig. 2. Disassembling spider silk fibers by ultra-sonication.** (**A**, **B**) Optical microscopy images of partially exfoliated silk fibers and fiber ends with exfoliated cores. (**C**) SEM image of a fully exfoliated fiber end. (**D**) SEM image of a partially exfoliated silk fiber. (**E**) AFM topography image of exfoliated nanofibrils. (**F**) Micro-fibril thinned out to the end by exfoliation of much smaller nanofibrils. Diameter distribution (**G**) and volume-weighted diameter distribution (**H**) of micro-fibrils found in Fig. 2C.

Interestingly, analysis of the AFM images suggested that the elongated sonication time did not further disintegrate the nanofibrils themselves. Furthermore, AFM images that were taken after 135 min (Fig. 3A) showed a much smaller amount of background material between the fibrils compared to AFM images taken after 30 min of sonication (Fig. 2E), where a significant amount of this non-fibrillar background material was present. This may seem counterintuitive, because one might assume that longer sonication time has the ability to break the fibrils into smaller fragments, thus leading to *more* non-fibrillar background. Our observation, however, can be explained by the fact that a natural *T. clavipes* MA fiber features an outer glycoprotein coating [20,55], which easily dissolves in water. (Assuming a 200 nm-thick [20,55] glycoprotein layer and a 4 μm-diameter fiber, the glycoprotein accounts for ≈20 vol% of the fiber.) These dissolved glycoproteins thus represent a much larger fraction of the solution protein in the early exfoliation process (Fig. 2E). Sonication for longer times then reduces the relative amount of unorganized background protein, as more nanofibrils from the fiber are exfoliated into the solution (Fig. 3A).

In essence, our exfoliation experiments suggest that, except for the soluble glycoprotein surface layer, virtually the entire *T. clavipes* MA silk fiber consists of ≈10 nm-diameter fibrils that were separated using relatively mild mechanical techniques. Since the non-fibrillar material of fully exfoliated fibers can be attributed to the glycoprotein fiber coating, we can conclude that the nanofibrils are the overwhelmingly dominant structural elements of these silk fibers. We did not observe any significant hierarchical organization of nanofibrils between their 10 nm size and the macroscopic silk fiber, i.e. there were no nanofibril bundles of a preferred size within the fiber. Instead, our SEM images of disintegrated fibers show branches with a wide range of diameters (Fig. 2, C, F, and H), possibly disconnected from the main fiber at random locations. The SEM evidence in Fig. 2C and 2D further shows that the separation of fiber components occurs strictly parallel to the fiber axis, suggesting that all fibrils are oriented parallel within the fiber. This is confirmed by AFM evidence of fiber fragments not completely exfoliated, where this parallel organization is revealed at the level of nanofibrils (inset in Fig. 3A). These observations are also in line with an NMR-based study conducted on *Trichonephila edulis* spider silk [54]. Based on these findings, we have developed the structural model of a *T. clavipes* MA fiber shown in Fig. 3B.

Notably, while the fiber completely disintegrated under ultrasound exposure, the nanofibrils themselves proved comparatively robust. This is in line with our recent study showing that silk of the Chilean recluse spider features highly anisotropic mechanical properties [56]. The work showed that the mechanical forces needed to separate recluse silk nanofibrils are five times smaller than the nanofibrils' tensile breaking forces. The work further suggested that van der Waals forces dominate interfibrillar binding. Corresponding conjectures regarding the internal mechanics of *T. clavipes* silk fibers are feasible. Facile ultrasound exfoliation of *T. clavipes* fibers would be in line with relatively weak van der Waals bonding between nanofibrils.

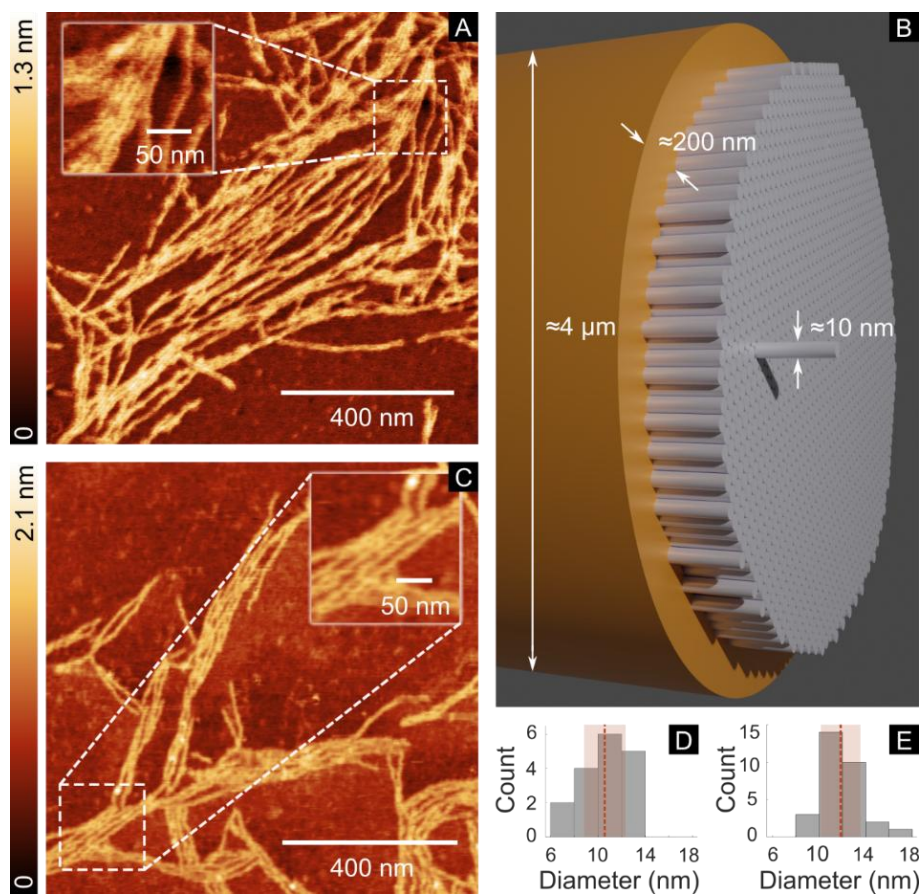

**Fig. 3. Exfoliated vs self-assembled spider silk nanofibrils and proposed structural model for *T. clavipes* major ampullate silk fiber.** (**A**) Exfoliated *T. clavipes* silk nanofibrils (AFM). (**B**) 3D-rendered illustration of the proposed nanofibrillar structure of *T. clavipes* silk fiber (thicknesses of fibrils and coating not drawn to scale for improved illustration). (**C**) Self-assembled silk nanofibrils (AFM). Diameter distribution of exfoliated (**D**) and self-assembled (**E**) nanofibrils. Average diameters and standard deviations are marked on the histograms in (D) and (E) by dashed lines and red shades, respectively. Insets in (A) and (C) indicate areas with aligned nanofibrils.

### 3.2 Spin-coating self-assembles native proteins into 10 nm nanofibrils

Having revealed that ≈10 nm-diameter fibrils, organized in a parallel fashion, are the fundamental building block of a *T. clavipes* MA fiber, the question of how these fibrils are formed and organized into a parallel bundle becomes very important, especially for the goal of making synthetic fibers inspired by spider silk. Building on some of our earlier observations on silkworm silk [38], we thus studied the process of nanofibril formation via self-assembly. We isolated major ampullate silk glands of *T. clavipes* spiders to extract their silk dope in native form as an aqueous gel. This gel was then diluted to 100 mg/L in deionized water and applied to freshly cleaved mica substrates via spin coating. This procedure reduces the $Na^+$ and $Cl^-$ concentrations, it reduces pH to ≈5.6 through adsorption of $CO_2$ followed by the production of carbonic acid, and it introduces shear. All of these conditions are known to also occur in the spider's spinneret at the onset of spinning and have been suggested to be triggers for spidroin self-assembly [47–49]. The obtained samples were imaged using dynamic-mode AFM to investigate the structural conformation of the protein. A typical AFM image is shown in Fig. 3C, revealing nanofibrils with a high degree of uniformity in terms of their diameters and heights. Interestingly, regarding their diameter, shape, and length, these self-assembled nanofibrils were very similar to the nanofibrils obtained via exfoliation of a native silk fiber from the same spider species (Figs. 2E and 3A).

### 3.3 The correspondence of self-assembled and exfoliated structures underlines the spidroin's tendency to nanofibrillate

Using quantitative analysis of the fibril diameters, we found both exfoliated nanofibrils and self-assembled nanofibrils featured diameters of (10±2) nm (Fig. 3, D and E). Furthermore, the self-assembled nanofibrils tended to aggregate side by side along the axial direction, just as observed in the exfoliated material. Examples for such bundles of parallel nanofibrils are magnified in the insets of Fig. 3A and 3C. We believe that the striking similarity between the nanofibrils produced via exfoliation (Fig. 3A) and self-assembly (Fig. 3C), as well as their tendency to align in a parallel fashion, is not a coincidence. Rather, we are convinced that these nanofibrils were formed based on the same underlying process, following an inherent capability implemented in spidroin. In the natural spinning process of the fiber, the silk dope is moved downstream in the silk gland, which induces significant shear, followed by removal of water [57] and ion exchanges [47–49]. It has been suggested that shear is a critical trigger for the formation of nanofibrils [58,59]. Similarly, in our spin-coating process, the silk solution is exposed to shear as the centripetal forces accelerate it radially away from the substrate, a process that gradually increases the silk concentration and ultimately removes liquid water entirely. While both processes share the presence of shear and removal of water, several other parameters may be significantly different, such as time-vs-concentration and time-vs-shear rate profiles. Nevertheless, the fact that they yield nanofibrils of almost identical morphologies suggests that the nanofibril formation mechanism in silk is relatively robust.

### 3.4 Spin-coating is applied to reveal the triggers of self-assembly

Based on the ability of our spin coating procedure to yield nanofibrils virtually identical to the natural spinning process, we can use this approach to learn more about the self-assembly of spidroin at the molecular scale. In a previous study, we found that the native silk protein of the silkworm *Bombyx mori* tends to open up and denature at low concentrations [39]. Therefore, we first explored the influence of protein concentration. We diluted the native protein concentration further, to 10 mg/L, using deionized water and spin-casted this solution onto mica sheets. Interestingly, the protein did not form long, straight nanofibrils, as observed at higher concentrations (Fig. 3C). Instead, the spidroin formed relatively loose and open structures (Fig. 4A) under these conditions, typically covering an area of 50 nm or more in diameter. The curly strands featured apparent widths of only 5–6 nm, about half of what we observed in the fibrils self-assembled at higher concentrations, and apparent heights of ≈0.2 nm. The true width of these strands is most likely significantly less because AFM generally overestimates the width of objects substantially smaller than the tip. We suspected that these curly strands represent denatured spidroin molecules that unfolded under low-concentration conditions. To further corroborate this hypothesis, we first estimated the lengths of the backbones of the fully unfolded single molecule of major ampullate silk proteins (MaSp). The silk of *T. clavipes* has been shown to feature two major components, MaSp-1 and MaSp-2 with ≈747 and ≈3,146 amino acids, which correspond to lengths of ≈260 nm and ≈1100 nm, respectively [60,61], assuming a length of 0.35 nm per amino acid [62]. Surprisingly, tracing the lengths of several of the curly strands in Figure 4a, we found good agreement with these lengths, supporting our hypothesis, and in line with our previous findings for the silkworm's native silk proteins [38]. Moreover, the dimensions of these structures have a surprising resemblance to what Parent et al. proposed as the hydrodynamic radius ($r_H$) of the major ampullate silk monomer of the black widow spider (*Latrodectus hesperus*) based on NMR experiments [43]. Based on their measured $r_H \approx 25$ nm, they illustrated the spidroin monomer as shown in Fig. 4B [43]. Our results directly visualize the actual spatial conformation of the spidroin in the low-concentration regime for the first time, and the similarity of the images thus provides support for the model developed by Parent et al. We note that AFM revealed slightly larger aggregate sizes. This could be caused by the AFM sample preparation process, in which the proteins are forced into a 2-dimensional (2D) conformation on the substrate, potentially increasing their lateral spread in comparison to the true 3D conformation in solution probed by NMR. Importantly, our results showing that spidroin denatures at low concentrations mean that it may lose the ability to self-assemble into nanofibrils under these conditions, and thus reveal the spidroin concentration as another important factor in the self-assembly process. Hence, our experimental method can also identify the threshold concentration of silk proteins needed for self-assembly.

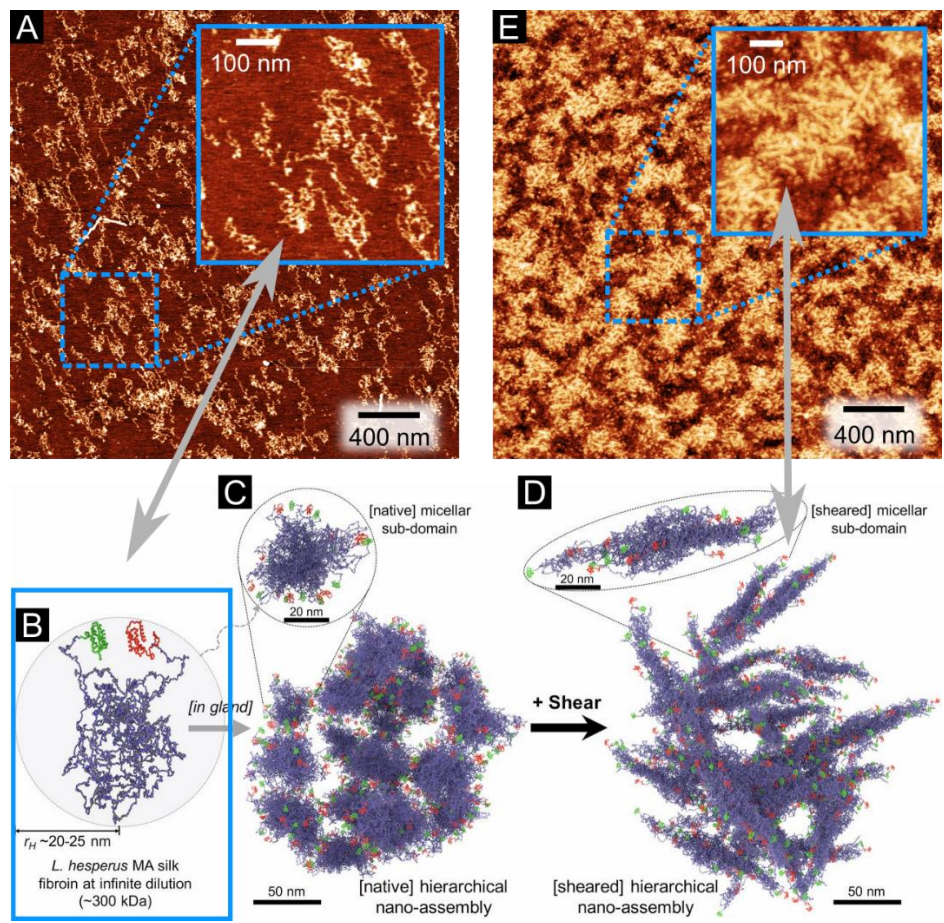

**Fig. 4. Comparison of NMR and cryo-TEM data with AFM images of self-assembled nanofibrils.** (**A**) Native *T. clavipes* spidroin in low (10 mg/L) concentration showing relatively loose and open structures (AFM, color scale: 0–0.4 nm). (**B**) Illustration of *L. hesperus* spidroin monomer based on the extrapolated hydrodynamic radius by NMR data. (**C**, **D**) Illustration of hierarchical organization of *L. hesperus* spidroin in at different conditions based on cryo-TEM imaging. (C) Native spidroin at high concentration. (D) Pre-fibrillar assemblies obtained by shearing the hierarchical micellar organizations. (**E**) *T. clavipes* spidroin at 1000 mg/L in PBS buffer (AFM, color scale: 0–5 nm). (B-D) Reproduced with permission [43]. 2018, United States National Academy of Sciences (United States).

Finally, we studied the influence of pH or concentration of $Na^+$ and $Cl^-$ ions on self-assembly, which have previously been suggested to be deciding factors in the spinning process [47–49]. Therefore, we diluted the native dope from the glands in PBS buffer at pH 7.4 rather than in deionized water. PBS buffer maintains a relatively high concentration of $Na^+$ and $Cl^-$ ions in the solution and keeps the pH neutral, akin to conditions in the storage sack of the spider. Spin-coated AFM samples made from such solutions at higher concentrations of 1000 mg/L no longer showed long, straight protein nanofibrils. Instead, the protein was organized in islands with 200–300 nm in diameter (Fig. 4E). The inset reveals that these islands are clusters of short protein fibril segments with diameters of ≈10 nm (similar to what we observed in unbuffered solutions, as shown in Fig. 3C) and relatively short lengths, in the range of 50–90 nm. The area between the islands appears to be composed of globules.

**3.5 Chemical and physical triggers during self-assembly**

The above experiment suggests that shear alone is not sufficient to trigger the full transformation of spidroin into long nanofibrils; instead, application of shear leaves the protein in a pre-assembled state, where a fraction is present as short, elongated, pre-fibrillar assemblies, without alignment. The structures observed in our AFM images are strikingly similar to the structures proposed by Parent et al. based on cryo-transmission electron microscopy (cryo-TEM) of *L. hesperus* spidroin (Fig. 4D) with respect to fibril diameters, fibril lengths, and size of assemblies. They diluted the samples in urea, which prevents the reduction of pH toward acidic conditions, and even though they applied shear to their solutions, they only observed transformation of the silk proteins from globular assemblies (Fig. 4C) into pre-fibrillar structures, but not complete assembly into long nanofibrils. Their observations are, thus, in close agreement with our observations and conclusions — that shear alone is not sufficient to trigger nanofibrillar assembly.

Our observation that the reduction of pH and $Na^+$ and $Cl^-$ concentrations are needed as well to trigger nanofibrillar assembly is in line with previous studies focusing on the functions of spidroin's N-terminal domain during self-assembly. One study used computational modeling to show that a high concentration of sodium chloride would hamper the dimerization of *Euprosthenops australis* spidroin's N-terminal domain, a key step to fibrillation [46], while another study revealed that neutral pH would also impede this step to avoid aggregation by utilizing recombinant "mini-spidroins" [63]. What our findings demonstrate is that the natural system uses several independent factors — concentration of $Na^+$ and $Cl^-$, pH, and shear — each of which is necessary to trigger the formation of nanofibrils. In the storage sack, none of the conditions necessary for assembly are met, which provides an effective mechanism to prevent premature assembly of the spidroin.

## 4. Conclusion

We have shown that ≈10 nm-diameter nanofibrils play a decisive role in the MA silk of the golden silk orb-weaver *T. clavipes*. We were able to virtually completely decompose natural spider silk fibers into such nanofibrils for the first time, as well as synthesize nanofibrils with almost identical morphology through self-assembly from native silk dope. We interpret this to mean that silk protein has a strong intrinsic mechanism to form nanofibrils of a defined morphology. To trigger this strong fibril-forming mechanism, several independent physical and chemical conditions — shear, reduction of pH, $Na^+$, and $Cl^-$ concentrations — need to be met simultaneously. Here, we have demonstrated for the first time that the absence of any of these triggers prevents premature assembly of spidroin in the storage sack and thus provides conditions enabling long-term storage. Importantly, there is a strong correlation between AFM, NMR, and cryo-TEM experiments in terms of the observed protein structures across a range of different experimental parameters (Fig. 4). This remarkable agreement of results from complementary methods suggests that these experiments observe the correct structure and that the information obtained can be combined toward a consensus model.

Notably, the ≈10 nm-diameter nanofibrils were the only structural element we observed throughout all exfoliation stages after the initial glycoprotein layer had dissolved. This shows that the entire core of the fiber consists of such nanofibrils, and we showed that they are organized in a parallel fashion. Moreover, our exfoliation results did not show evidence of any systematic hierarchical organization among nanofibrils, i.e., we did not see nanofibril bundles of distinct diameters others had suggested [13,53,54]. We used our evidence to develop a refined structural model of *T. clavipes* spider silk fibers (Fig. 3B).

Our work has shown two complementary nanofibril preparation pathways, disassembly of micro-sized fibers (top-down) vs. self-assembly of molecules (bottom–up), to make spider silk in this form for the first time, with great potential for new applications [25–27,30]. In addition, the methods we introduced provide new ways to study the role of nanofibrils, their formation, and organization. A limitation of our method, the combination of spin coating with AFM, is that it does not fully replicate the natural spinning system of spiders. In the natural system, the viscous spidroin solution self-assembles into nanofibrils and ultimately transforms into a microscopic fiber. Although our *in-vitro* approach has the capability to study the parameters of nanofibril self-assembly, it is not yet developed to create microscopic fibers as the end product.

However, our method unlocks a territory to study the most important stage of the fiber formation, which is still a missing link in the research of silk synthesis, i.e., the conditions of nanofibrils formation. Currently, most synthetic silks are inferior to their natural counterparts in terms of their mechanical properties [47,64–66]. We believe that this is because the structure of the natural silk fiber, including the critically important nanofibrils, has not been fully reproduced, yet [65,66]. Our system provides a relatively easy-to-implement yet versatile *in-vitro* approach to evaluate the influence of different physical and chemical parameters on self-assembly. It can be used to optimize the spinning parameters of synthetic silks made from recombinant silk proteins to yield fibers structurally closer to natural silk. We believe our findings will have a significant impact on ongoing efforts to mass-produce desirable high-performance materials inspired by spider silk.

**Acknowledgments**

Funding**:** This work was supported by the National Science Foundation, grants DMR-1352542 and DMR-1905902.

The authors also thank William & Mary Applied Research Center (ARC) for permitting access to their equipment.

**Author contributions:**
Exfoliation experiments: DP and LL
Self-assembly experiments: QW, CW, and DP
Data analysis and interpretation (exfoliation): DP, LL, and HCS
Data analysis and interpretation (self-assembly): QW, CW, DP, and HCS
Writing: DP, LL, and HCS
Conception and supervision: HCS

**Competing interests:** No conflict of interest has been declared by the authors.

**Data and materials availability:** The data supporting the findings of this study are available free of charge at https://doi.org/10.7910/DVN/P91YZ8 [51].